\documentclass[aps,prb,amsmath,amssymb,twocolumn,showpacs,superscriptaddress,final,floatfix]{revtex4-1}
\usepackage{graphicx}	
\usepackage{color}	
\usepackage[normalem]{ulem}
\usepackage{hyperref}
\hypersetup{
  colorlinks,
  citecolor=blue,
  linkcolor=blue,
  urlcolor=blue}

\newcommand{\BFA}{BaFe$_2$As$_2$}
\newcommand{\CaFA}{CaFe$_2$As$_2$}
\newcommand{\CsFA}{CsFe$_2$As$_2$}

\newcommand{\ie}{{\em i.e.}}

\begin{document}

\title{Low work function in the 122-family of iron-based superconductors}

\author{H. Pfau}
\email{hpfau@lbl.gov}
\affiliation{Advanced Light Source, Lawrence Berkeley National Laboratory, Berkeley, CA 94720, USA}
\affiliation{Stanford Institute for Materials and Energy Sciences, SLAC National Accelerator Laboratory, Menlo Park, CA 94025, USA}
\author{H. Soifer}
\affiliation{Stanford Institute for Materials and Energy Sciences, SLAC National Accelerator Laboratory, Menlo Park, CA 94025, USA}
\affiliation{Raymond and Beverly Sackler School of Physics and Astronomy, Tel Aviv University, Tel Aviv 4R73+8Q, Israel}
\author{J.~A. Sobota}
\affiliation{Stanford Institute for Materials and Energy Sciences, SLAC National Accelerator Laboratory, Menlo Park, CA 94025, USA}
\affiliation{Geballe Laboratory for Advanced Materials, Department of Applied Physics, Stanford University, Stanford, CA 94305, USA}
\author{A. Gauthier}
\affiliation{Stanford Institute for Materials and Energy Sciences, SLAC National Accelerator Laboratory, Menlo Park, CA 94025, USA}
\affiliation{Geballe Laboratory for Advanced Materials, Department of Applied Physics, Stanford University, Stanford, CA 94305, USA}
\author{C. R. Rotundu}
\affiliation{Stanford Institute for Materials and Energy Sciences, SLAC National Accelerator Laboratory, Menlo Park, CA 94025, USA}
\author{J. C. Palmstrom}
\affiliation{Stanford Institute for Materials and Energy Sciences, SLAC National Accelerator Laboratory, Menlo Park, CA 94025, USA}
\affiliation{Geballe Laboratory for Advanced Materials, Department of Applied Physics, Stanford University, Stanford, CA 94305, USA}
\author{I.~R.~Fisher}
\affiliation{Stanford Institute for Materials and Energy Sciences, SLAC National Accelerator Laboratory, Menlo Park, CA 94025, USA}
\affiliation{Geballe Laboratory for Advanced Materials, Department of Applied Physics, Stanford University, Stanford, CA 94305, USA}
\author{G.-Y. Chen}
\author{H.-H. Wen}
\affiliation{Center for Superconducting Physics and Materials, National Laboratory of Solid State Microstructures and Department of Physics, National Center of Microstructures and Quantum Manipulation, Nanjing University, Nanjing 210093, China}
\author{Z.-X. Shen}
\affiliation{Stanford Institute for Materials and Energy Sciences, SLAC National Accelerator Laboratory, Menlo Park, CA 94025, USA}
\affiliation{Geballe Laboratory for Advanced Materials, Department of Applied Physics, Stanford University, Stanford, CA 94305, USA}
\affiliation{Department of Physics, Stanford University, Stanford, California, USA}
\author{P. S. Kirchmann}
\email{kirchman@stanford.edu}
\affiliation{Stanford Institute for Materials and Energy Sciences, SLAC National Accelerator Laboratory, Menlo Park, CA 94025, USA}

\date{\today}


\begin{abstract}

We determine the work functions of the iron arsenic compounds $A$Fe$_2$As$_2$ ($A=\mathrm{Ca, Ba, Cs}$) using photoemission spectroscopy to be 2.7\,eV for \CaFA, 1.8\,eV for \BFA, and 1.3\,eV for \CsFA. The work functions of these 122 iron-based superconductors track those of the elementary metal $A$ but are substantially smaller. The most likely explanation of this observation is that the cleaving surface exposes only half an $A$-layer. The low work function and good photoemission cross section of \BFA~and \CsFA~enable photoemission even from a common white LED light.

\end{abstract}

\maketitle

\section{Introduction}

Iron-based high-temperature superconductors (FeSCs) are an intensely studied class of quantum materials. Superconductivity usually appears when nematic and antiferromagnetic phases are suppressed by doping or substitution \cite{paglione_2010}. A prime example are the phase diagrams of doped \BFA, which belongs to the 122 family \cite{shibauchi_2014}. The electronic structure is complex with several bands crossing the Fermi level. Angle-resolved photoemission spectroscopy (ARPES) is an indispensable probe of the multi-band and multi-orbital character \cite{yi_2017}. Nematicity and magnetism lead to strong modifications of the band structure \cite{hsieh_2008,yang_2009,kim_2011,wang_2013,kondo_2010,liu_2009_prb,jensen_2011,yi_2009_prb,de_jong_2010,liu_2010_natphys,richard_2010,shimojima_2010,zabolotnyy_2009,yi_2011_pnas,pfau_2019_prb, fedorov_arxiv,watson_2019}. Further photoemission studies with higher resolution and in combination with different tuning parameters will be needed to form a consensus on band assignment and energy scales of these two ordering phenomena \cite{yi_2011_pnas,pfau_2019_prl, fedorov_arxiv,watson_2019}.

The interpretation of photoemission spectra of FeSCs is complicated by the appearance of surface-related bands \cite{jensen_2011,pfau_2019_prb,heumen_2011,watson_2019}. The distinction between surface and bulk related features is indispensable for a correct band assignment. The dispersion of surface bands depends on the surface termination and reconstruction \cite{heumen_2011}. However, there is no consensus in the literature on this issue. The work function is an excellent indicator of the surface termination.

In a more general context, accurate analysis of photoemission spectra requires knowledge of the sample work function $\Phi$ because it influences the kinetic energy and the emission angle of the outgoing electrons. The influence of the work function is particularly pronounced when the photon energy is of the order of the work function such as in laser-based ARPES \cite{fero_2014}. It is therefore valuable to have a reliable information on the work function. 

Previously, the work function of a few 122 FeSCs was estimated by measuring the barrier height in scanning tunneling spectroscopy: \BFA~with $\Phi \sim1.5$\,eV, \CaFA~with $\Phi \sim1.9$\,eV \cite{massee_diss,hoffmann_2011} and Pr$_{x}$Ca$_{1-x}$Fe$_2$As$_2$ with 4.5\,eV for the $1\times 1$ surface termination and 3.6\,eV for the $2 \times 1$ reconstructed surface \cite{zeljkovic_2013}. The first two values are unusually low. The work function of most metals lies between 4 and 5\,eV \cite{crc_handbook}. We expect the low work function in 122 FeSCs to influence laser-based ARPES more severely than in other compounds. 

Here, we determine the work function of three $A$Fe$_2$As$_2$ FeSCs with $A=\mathrm{(Ca,Ba,Cs)}$ using laser-based photoemission spectroscopy \cite{hufner_book}. The work function for freshly cleaved \CaFA~is 2.7\,eV, 1.8\,eV for \BFA, and 1.3\,eV for \CsFA. These values track the work functions of the corresponding alkali and alkaline earth metals, but are considerably smaller in the 122 FeSCs. While there is no consensus in the literature which atoms terminate the cleaving surface, we explain our observation with a surface structure containing only half the $A$-layer. The large distance between the surface atoms leads to smoothing of the charge distribution, which in turn lowers the work function. We observe that the work function changes considerably over the course of hours in ultra-high vacuum conditions due to adsorption of residual gas molecules onto the surface. In particular, the work function in \CsFA~is one of the lowest reported work functions for any material.

\section{Methods}

\begin{figure}
\begin{center}
\includegraphics[width=\columnwidth]{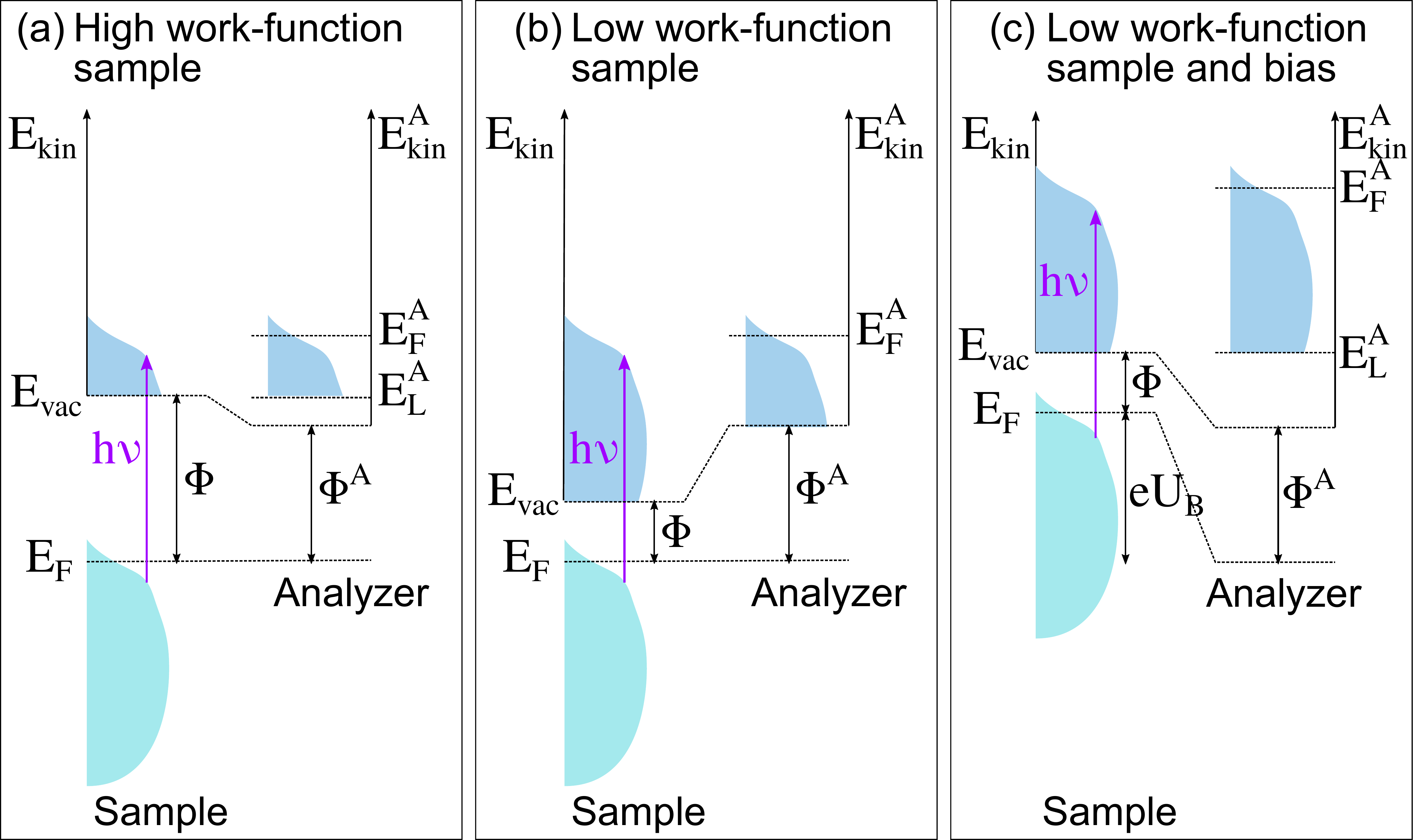}
\caption{Energy diagram for photoemission (a) on a sample with a high work function, \ie~ $\Phi>\Phi^\mathrm{A}$; (b) on a sample with a low work function $\Phi<\Phi^\mathrm{A}$; and (c) on a low work-function sample with a negative bias voltage $U_\mathrm{B}$ to overcome the vacuum energy barrier shown in (b).}
\label{Fig:energy_dia}
\end{center}
\end{figure}

Figure \ref{Fig:energy_dia} shows energy diagrams for photoemission from samples with different work functions. $\Phi$ is defined as the energy required to promote an electron from the Fermi level $E_\mathrm{F}$ to the vacuum level $E_\mathrm{vac}$. According to Fig. \ref{Fig:energy_dia}(a), photons with energy $h\nu$, that is larger than the sample work function $\Phi$, emit electrons with a kinetic energy 
\begin{equation}
 E_\mathrm{kin} = h\nu - \Phi - (E_i - E_\mathrm{F})
\end{equation}
$(E_i - E_\mathrm{F})$ is the electron binding energy with respect to the Fermi energy $E_\mathrm{F}$. The sample and the analyzer are electrically connected and therefore share the same Fermi energy $E_\mathrm{F}$. The difference of their work functions $\Phi$ and $\Phi^\mathrm{A}$ then changes the kinetic energy of the photoemitted electrons such that the measured kinetic energy $E^\mathrm{A}_\mathrm{kin}$ is given by
\begin{equation}
 E^A_\mathrm{kin} = h\nu - \Phi^A - (E_i - E_\mathrm{F})
\end{equation}
where $\Phi^A$ is the work function of the material that covers the entrance of the analyzer.
Photoemitted electrons with zero kinetic energy $E_\mathrm{kin}$ are accelerated towards the entrance of the analyzer and appear at non-zero kinetic energy $E^A_\mathrm{kin}$ for $\Phi>\Phi^\mathrm{A}$. This low-energy cut-off forms a parabola as function of in-plane momentum as seen for example in Fig.~\ref{Fig:bias_dep}(e). The minimum of this parabola $E_\mathrm{L}^A$ appears at normal emission, \ie~zero in-plane momentum.. The sample work function $\Phi$ can then be determined by
\begin{equation}
 \Phi = h\nu - (E^\mathrm{A}_\mathrm{F} - E^\mathrm{A}_\mathrm{L})
 \label{eqn:workfunction}
\end{equation}
where $E^\mathrm{A}_\mathrm{F}$ is the Fermi level as measured in the analyzer.

Figure \ref{Fig:energy_dia}(b) illustrates that electrons with a small kinetic energy do not reach the analyzer and hence $E^\mathrm{A}_\mathrm{L}$ is not accessible when the work function of the sample is lower than the work function of the analyzer. We overcome this problem by applying a negative bias voltage to the sample which accelerates electrons towards the analyzer and leads to an energy diagram as shown in Fig.~\ref{Fig:energy_dia}(c). For the sake of simplicity we report here the absolute value $U_\mathrm{B}$ of the bias voltage.

Our experimental setup is based on a Coherent RegA 9040 amplifier providing a fundamental photon energy of 1.50\,eV with a repetition rate of 312\,kHz. 6\,eV photons for photoemission are generated by quadrupling the amplifier output. We detect the photoemitted electrons with a Scienta R4000 hemispherical analyzer. The overall energy resolution is 50\,meV and dominated by the bandwidth of the ultrafast 6\,eV pulses. Samples are cleaved {\it{in-situ}} at a base pressure of $1\times10^{-10}$\,Torr. All measurements were performed at room temperature unless otherwise noted. We applied a variable bias voltage using standard alkaline batteries. We studied single crystals of \CaFA, \BFA, and \CsFA. They were grown from an FeAs flux as described previously \cite{chu_2009,rotundu_2010}.

\section{Results and Discussion}

\begin{figure}
\begin{center}
\includegraphics[width=\columnwidth]{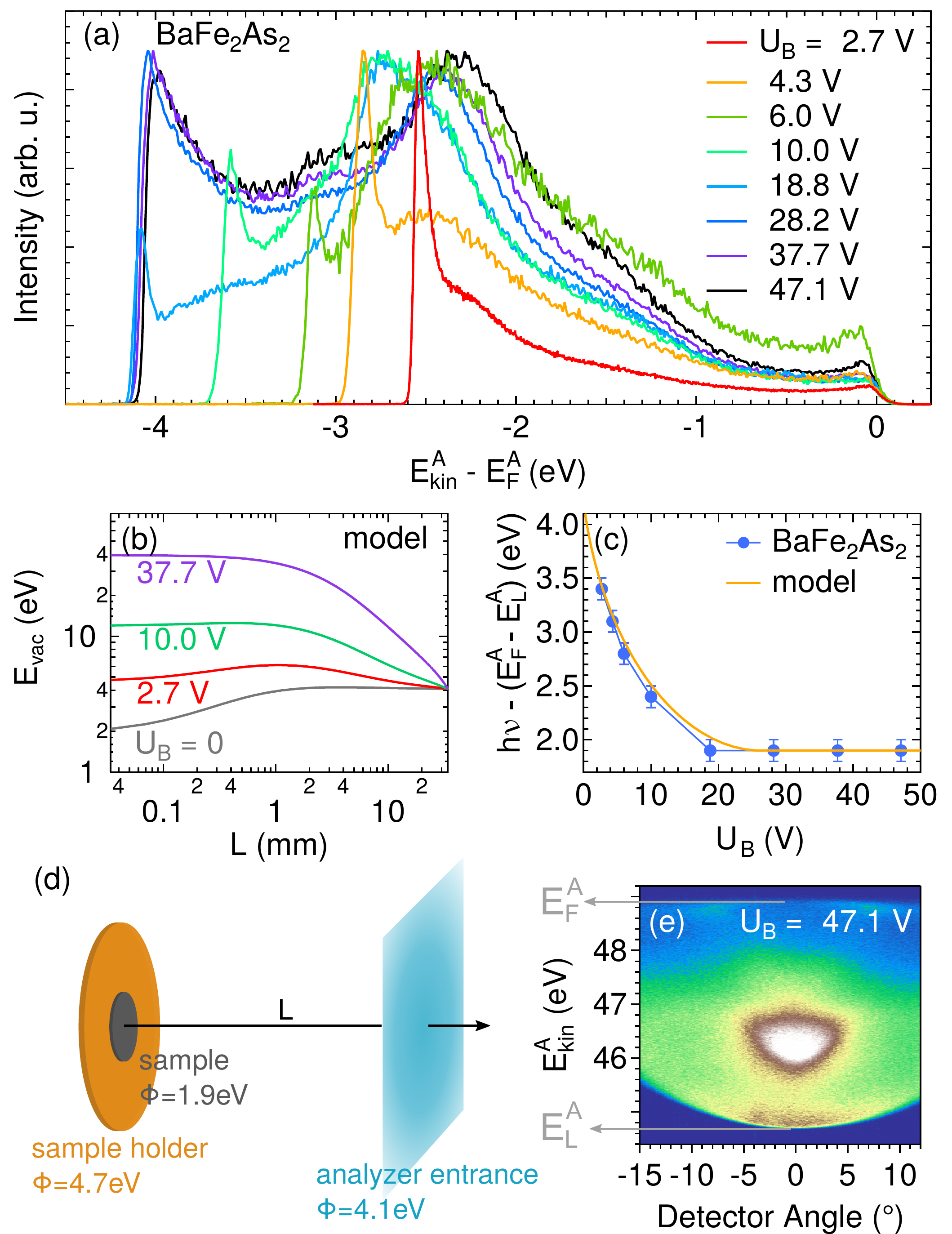}
\caption{Photoemission spectra of \BFA~as function of bias voltage $U_\mathrm{B}$. (a) EDCs at normal emission as a function of $U_\mathrm{B}$. (b) Electrostatic potential $E_\mathrm{vac}$ calculated from the model displayed in (d) for different $U_\mathrm{B}$. (c) Apparent work function $h\nu - (E^\mathrm{A}_\mathrm{F} - E^\mathrm{A}_\mathrm{L})$ derived from the EDCs shown in (a) (dots) and calculated using the model in (d) (line). (d) Schematics of the electrostatic model. (e) Typical photoemission spectrum at high bias $U_\mathrm{B}$.
}
\label{Fig:bias_dep}
\end{center}
\end{figure}

Figure \ref{Fig:bias_dep}(a) presents energy distribution curves (EDCs) of \BFA~at normal emission for different bias voltages $U_\mathrm{B}$. We observe that the width of the spectrum $(E^\mathrm{A}_\mathrm{F} - E^\mathrm{A}_\mathrm{L})$ increases with increasing $U_\mathrm{B}$. While this behavior is expected for small bias voltages $U_\mathrm{B}<(\Phi^\mathrm{A}-\Phi)$ from our considerations in Fig.~\ref{Fig:energy_dia}(b,c), it is surprising for larger bias $U_\mathrm{B}>(\Phi^\mathrm{A}-\Phi)$. We plot the apparent work function $h\nu - (E^\mathrm{A}_\mathrm{F} - E^\mathrm{A}_\mathrm{L})$ in Fig.~\ref{Fig:bias_dep}(c). It approaches a constant value of 1.9\,eV for $U_\mathrm{B}>30$\,eV. 

We explain the bias dependence by considering an electrostatic model as shown in Fig.~\ref{Fig:bias_dep}(d). We assume the sample to be a disk with a radius of 0.5\,mm, which approximates our sample size.  We assume a work function of $\Phi=1.9$\,eV for the sample as obtained from $h\nu - (E^\mathrm{A}_\mathrm{F} - E^\mathrm{A}_\mathrm{L})$ for large $U_\mathrm{B}$. The shape of the sample holder is approximated by a disk with a radius of 5\,mm. It consists of copper with a work function of 4.7\,eV. The analyzer entrance is considered to be an infinite plate 34\,mm away from the sample, which corresponds to the distance in our experiment. The graphite-coated analyzer has a work function of 4.1\,eV. 

Using this model, we calculate the electrostatic potential $E_\mathrm{vac}$ along the axis $L$ ranging from the sample to the analyzer, which we plot in Fig.~\ref{Fig:bias_dep}(b). The result demonstrates that photoemitted electrons with a small kinetic energy are not able to reach the detector if the bias voltage is too small to overcome the local potential well near the sample surface. The apparent work function $h\nu - (E^\mathrm{A}_\mathrm{F} - E^\mathrm{A}_\mathrm{L})$ from this model plotted in Fig.~\ref{Fig:bias_dep}(c) reproduces the experimentally determined data well. Our model shows that a large enough bias needs to be applied to correctly extract the work function $\Phi$ from the spectral width $(E^\mathrm{A}_\mathrm{F} - E^\mathrm{A}_\mathrm{L})$. We made sure that this width is independent of $U_\mathrm{B}$ for the following measurements of $\Phi$. We show a typical ARPES spectrum in Fig.~\ref{Fig:bias_dep}(e) where $E^\mathrm{A}_\mathrm{F}$ is seen at large kinetic energies and the parabola of the low-energy cut-off at low kinetic energies.

\begin{figure}
\begin{center}
\includegraphics[width=\columnwidth]{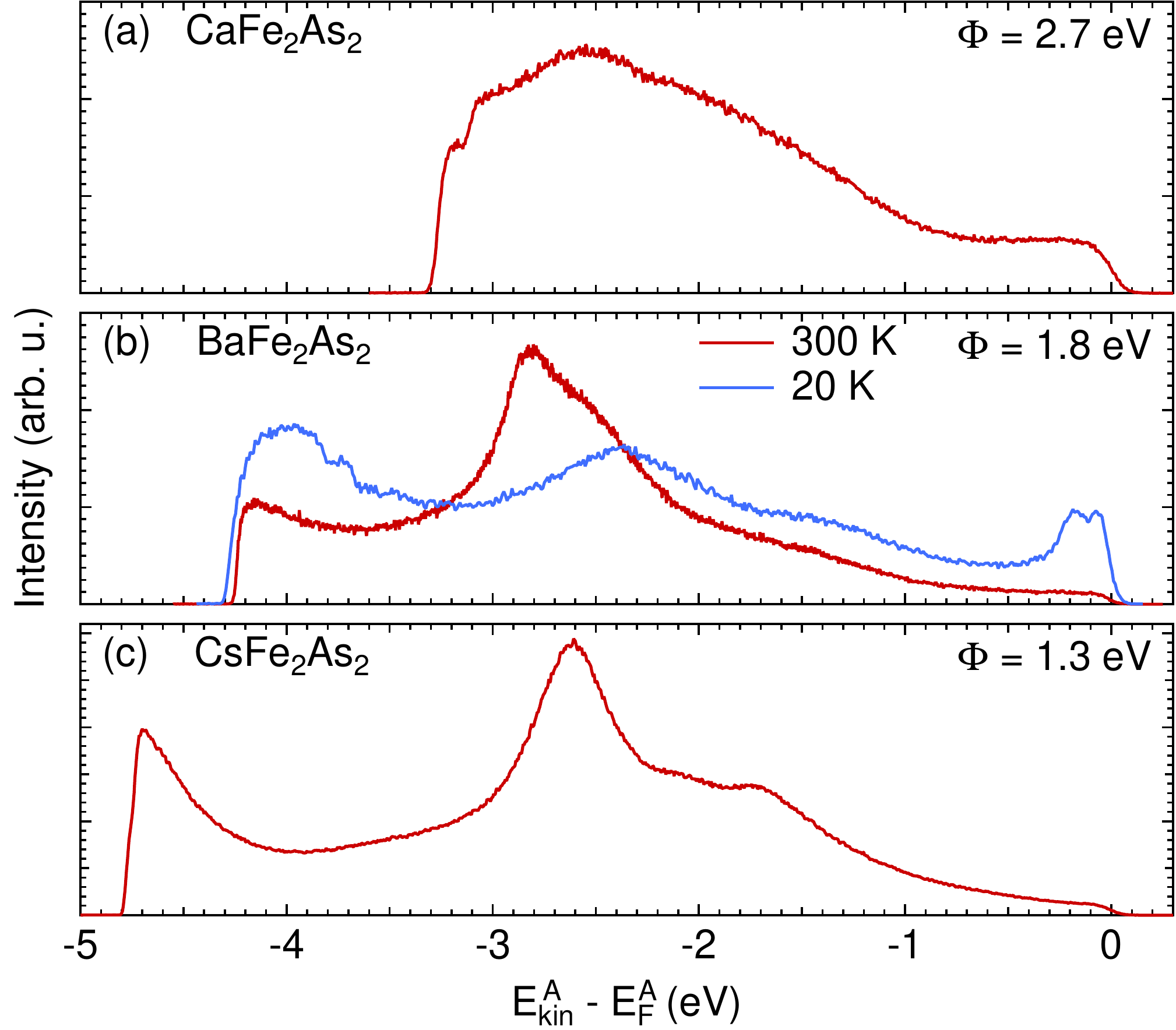}
\caption{EDCs at normal emission of (a) \CaFA~($U_\mathrm{B}=28.2$\,V), (b) \BFA~($U_\mathrm{B}=47.1$\,V), and (c) \CsFA~($U_\mathrm{B} = 37.4$\,V). We show additional data on a cleave at 20\,K for \BFA~($U_\mathrm{B} = 36.3$\,V). The work function $\Phi$ is determined from the position of the Fermi level $E^\mathrm{A}_\mathrm{F}$ and of the low-energy cut-off $E^\mathrm{A}_\mathrm{L}$ from the data at 300\,K applying Equation \ref{eqn:workfunction}.
}
\label{Fig:workfunction}
\end{center}
\end{figure}

\begin{table}
\centering
\begin{tabular}{c c | c c c }
\toprule
& $\Phi$ (eV) & & $\Phi_\mathrm{Alk}$ (eV) & $n \mathrm{(nm^{-3})}$ \\
\hline
\CaFA & 2.7 & Ca & 2.9 & 45.8\\
\BFA & 1.8 & Ba & 2.5 & 31.6\\
\CsFA & 1.3 & Cs & 2.0 & 8.6\\
\hline
&& As & 3.8 & \\
\toprule 
\end{tabular}
\caption{
Work function $\Phi$ of the three studied 122 FeSCs The error of the measurement is 0.1\,eV. We compare these values to the work function $\Phi_\mathrm{Alk}$ of the corresponding polycrystalline elemental alkali and alkaline metals \cite{crc_handbook} as well as the valence electron density $n$ of the elemental metals \cite{crc_handbook}. We added the work function of As \cite{crc_handbook}.
}
\label{tab:workfunction}
\end{table}
\begin{figure}
\begin{center}
\includegraphics[width=\columnwidth]{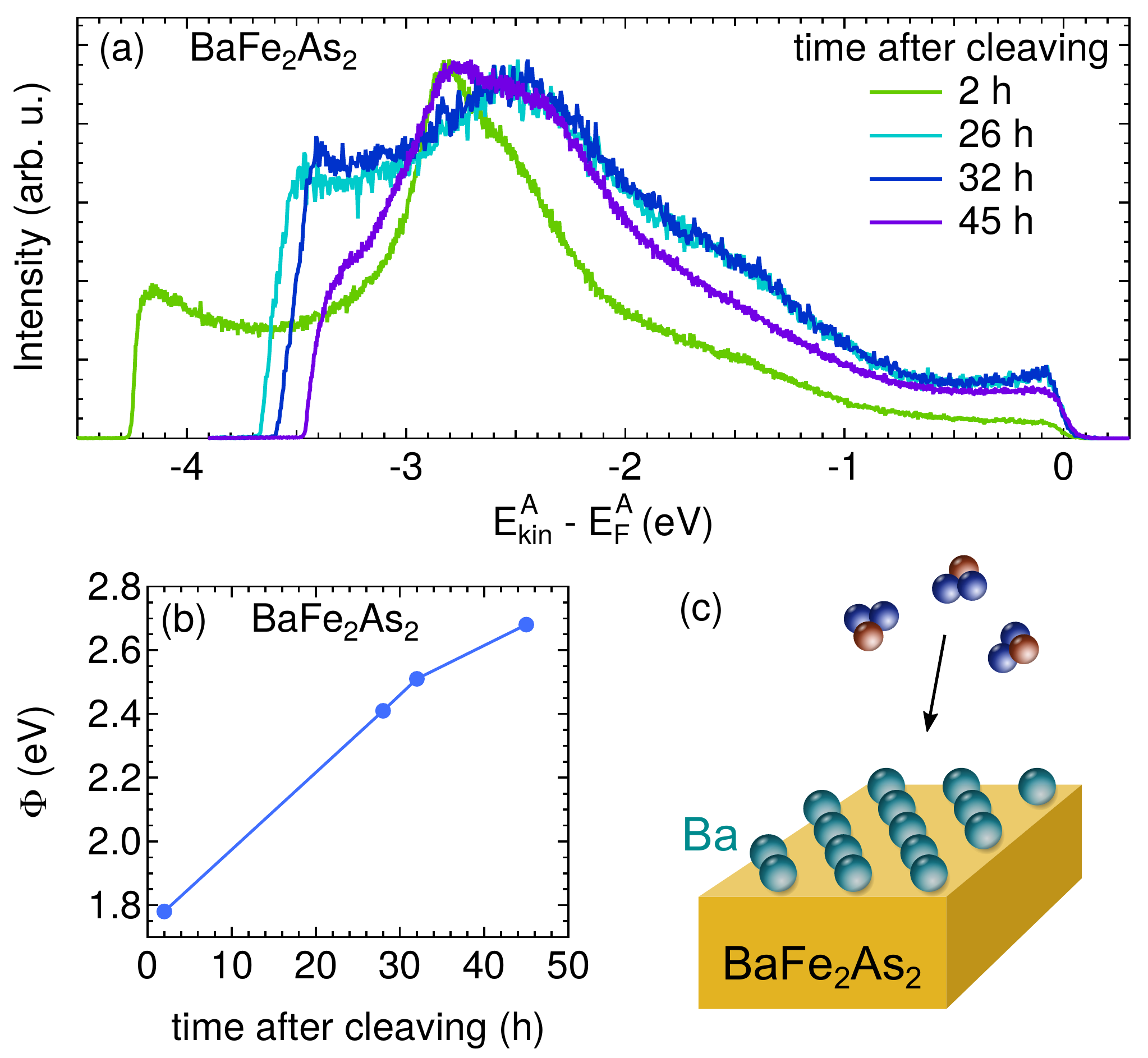}
\caption{Time-dependence of the work function. (a) EDCs at normal emission as function of time after cleaving and (b) associated change of $\Phi$. The data for 2\,h after cleaving were determined on a second sample. The general trend of $\Phi$ as function of time after cleaving was reproduced for a number of cleaves.
}
\label{Fig:time_dep}
\end{center}
\end{figure}

We determined the work function of three members of the 122 FeSC family: \CaFA, \BFA, and \CsFA. The EDCs at normal emission are shown in Fig.~\ref{Fig:workfunction}. All three data sets were obtained immediately after cleaving. We summarize the results in Table \ref{tab:workfunction}. The error of the measurement determined from the width of the low-energy cut-off as well as the time and sample dependence is 0.1\,eV. The work function of all three samples is much lower than the work function of most metals, for example $5.31$\,eV for Au(111) or 4.81\,eV for Fe(111) \cite{crc_handbook}. The work functions we determined on \BFA~and \CaFA~agree reasonably well with those reported by Massee \cite{massee_diss} and are only slightly higher. 

Cleaving $A$Fe$_2$As$_2$ is expected to leave the FeAs bond intact and to expose either As or $A$-atoms \cite{hoffmann_2011}. Tunneling images of cleaved surfaces typically show only half the atoms expected for a full $A$ or As layer, mostly in a $\sqrt{2}\times\sqrt{2}$ or a $2\times 1$ reconstruction \cite{hoffmann_2011}. However, there is no consensus in literature if the As or the $A$ atoms terminate the cleaving surface. Currently, proposals for the most common surface reconstructions include (i) a $\sqrt{2}\times\sqrt{2}$ or $2\times 1$ reconstructed half layer of $A$ atoms \cite{hsieh_2008,yin_2009,massee_2009,zhang_prb_2010,heumen_2011,nishizaki_2011,gao_2010,zeljkovic_2013,song_2013,li_2014_prl}, (ii) a full layer of As with specific tunneling matrix elements \cite{nascimento_2009,li_prb_2012,niestemski_2009_arxiv,nishizaki_2011} or with a $2\times 1$ dimerization\cite{li_2019_prb,huang_2013}, or (iii) a complete $A$ layer with a $\sqrt{2}\times\sqrt{2}$ buckling reconstruction \cite{li_2019_prb}. While Massee reported that the work function is independent of the surface reconstruction and the same over dozens of cleaved samples \cite{massee_diss}, Zeljkovic found different work functions for differently reconstructed surfaces \cite{zeljkovic_2013}. With the finite beam spot size, we likely probe different surface reconstructions at the same time but can only detect the lowest work function.

In Table \ref{tab:workfunction}, we compare $\Phi$ to the work function $\Phi_\mathrm{Alk}$ of the elements Ca, Ba, and Cs, and of As. $\Phi$ is consistently lower than the work function of the corresponding elements. However, we observe a clear correlation between $\Phi_\mathrm{Alk}$ and $\Phi$. This correlation suggests that the surfaces of cleaved $A$Fe$_2$As$_2$ are consistently terminated by $A$ atoms. 

Generally, the work function depends on three energy scales: (1) The chemical potential of the bulk, (2) a surface dipole created by the electronic wave function spilling out into the vacuum, and (3) an opposing surface dipole created by the smoothing of the charge distribution on a rough surface \cite{lang_1971,smoluchowski_1941,wigner_1935}. (1) and (2) are governed by the valence electron density $n$, which we list in Table \ref{tab:workfunction} for the elementary metals. The work function is expected to decrease with decreasing $n$ \cite{lang_1971}, which agrees with the trends of $\Phi_\mathrm{Alk}$ and $\Phi$ of the FeSCs presented in Table \ref{tab:workfunction}. The difference between $\Phi$ and $\Phi_\mathrm{Alk}$ can thus be attributed to (3). In general, a rougher surface has a larger dipole due to the smoothing of the electron distribution, which reduces the work function \cite{smoluchowski_1941}. A half layer of $A$ atoms on the surface results in a substantially rougher surface than for a complete layer and can explain why the work function in 122 FeSCs is smaller than in the polycrystalline metals. A similar but smaller effect can be expected for a buckling reconstruction of the $A$ atoms.

\begin{figure}
\begin{center}
\includegraphics[width=\columnwidth]{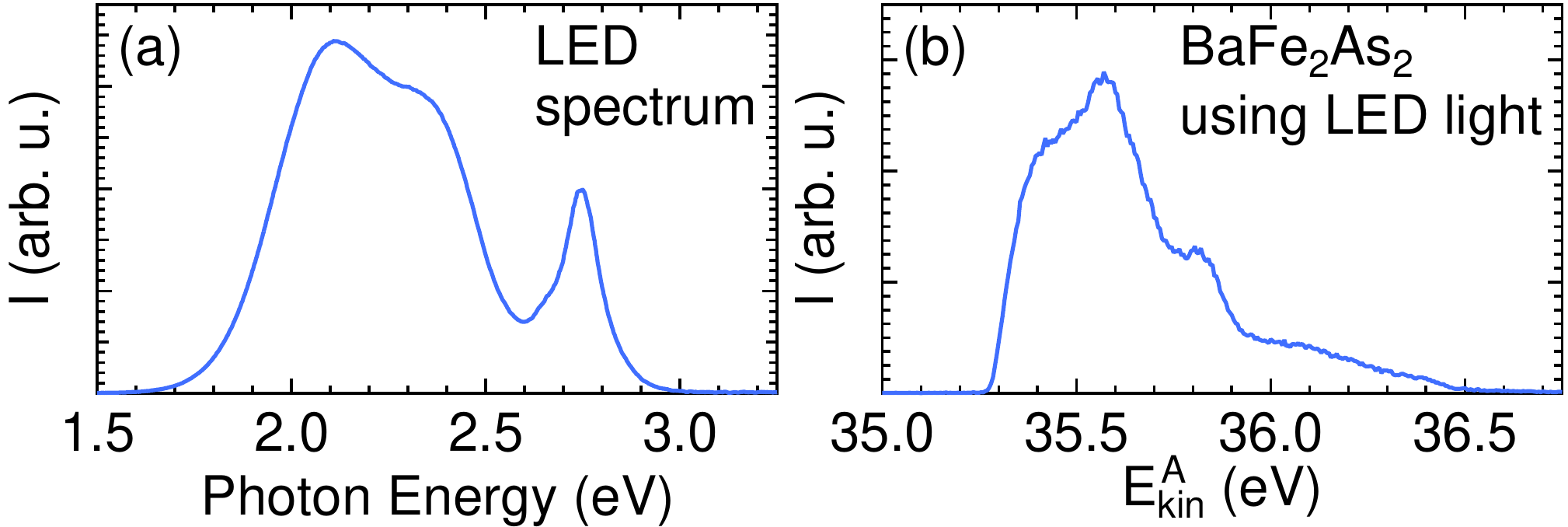}
\caption{Photoemission using LED light. (a) Optical spectrum of common white LED after transmission through a BK7 window. (b) Angle-integrated photoemission spectrum of \BFA~using LED light and a bias of $U_\mathrm{B} = 37.6$\,V.}
\label{Fig:LED}
\end{center}
\end{figure}

It has been shown that the surface reconstructions depend on cleaving temperature and can undergo structural transitions as function of temperature \cite{massee_2009,massee_diss,hsieh_2008}. Our measurements of $\Phi$ in \BFA~do not show a significant difference between cleaving at room temperature and 20\,K, see Fig.~\ref{Fig:workfunction}(b). We conclude that the main mechanism of the low work function is the half layer at the surface and not its particular ordering. This conclusion is supported by our observation (not shown) that an increase in temperature from 20\,K to 200\,K across the reported structural transition of the surface reconstruction does not lead to a sudden change in work function.

For photoemission studies of 122 FeSCs, it is important to understand that the work function changes over time due to the adsorption of residual gas molecules. We therefore studied the time dependence of the work function on \BFA~at room temperature at a pressure of $1\times10^{-10}$\,Torr and present the results in Fig.~\ref{Fig:time_dep}. We find that the work function increases by almost 1\,eV over the course of 48\,h. We explain this relatively large change by the difference between the small initial value of $\Phi$ and the work function of $\sim 4$\,eV of typical adsorbates such as water.

The low work function of 122 FeSCs can cause unexpected experimental challenges. Remarkably, the LED light that illuminates the inside of the vacuum chamber leads to substantial photoemission intensities from the \BFA~sample. In Fig.~\ref{Fig:LED}(a), we present the optical spectrum of the LED after transmission through a BK7 window as used as viewports in our vacuum system. The main spectral intensity is located between 2 and 2.5\,eV photon energy, which is large enough to overcome the work function of \CsFA~and \BFA. A bias is again necessary to overcome the potential barrier and detect the photoelectrons with the analyzer as we showed earlier. The corresponding photoemission spectrum of \BFA~is plotted in Fig.~\ref{Fig:LED}(b). Due to the considerable cross section for photoemission from these white LED lights, we performed our experiments with the lights switched off. 

The work function of \CsFA~is one of the lowest reported for any material. Coating a surface with Cs has been widely used to lower a sample's work function and values between 1.0\,eV and 1.4\,eV are typical \cite{yuan_2016,su_1983,bazarov_2008,karkare_2014,zhang_2011_apl,desplat_1980,sun_2007}. This technique is for example applied for photocathodes such as Cs-O coated GaAs \cite{su_1983,bazarov_2008,karkare_2014,zhang_2011_apl}. Low work function materials are generally desirable for laser driven electron sources, because light in the visible range can be used. Such lasers typically provide a larger photon flux than UV sources.


\section{Summary}
The work function of 122 FeSCs determined by photoemission spectroscopy is lower than in other quantum materials. In particular, 1.3\,eV found for \CsFA~is one of the lowest reported work functions for any material. The work function correlates with that of the alkaline earth and alkali atoms present at the surface, but is lowered by the roughness of the surface that contains only half an atomic layer. The low work function and good photoemission cross section leads to photoemission from a white LED light. We demonstrated that for low work function materials photoelectrons with a low kinetic energy are not able reach the analyzer and require the application of a bias voltage of up to 40\,V to overcome the local potential well. The work function changes considerably over time under ultrahigh vacuum conditions due to the adsorption of residual gas molecules onto the sample surface.


\begin{acknowledgments}
H.P. acknowledges support from the Alexander von Humboldt Foundation and from the German Science Foundation (DFG) under reference PF 947/1-1. J.C.P. is supported by a Gabilan Stanford Graduate Fellowship and a NSF Graduate Research Fellowship (Grant No. DGE-114747). The work at the ALS is supported by US DOE under contract no. DE-AC02-05CH11231. This work was supported by the U.S. Department of Energy, Office of Science, Basic Energy Sciences, Materials Sciences, and Engineering Division, under Contract No. DE-AC02-76SF00515.
\end{acknowledgments}


\bibliography{work_function_manuscript}

\end{document}